# Broad Brush Cosmos

Chris L. Carilli (NRAO)

*An innovative approach to map the large-scale structure in the Universe sidesteps the conventional need to observe millions of galaxies individually, and holds promise for both astrophysical and cosmological studies.*

The study of large-scale structure in the Universe — the distribution of galaxies on scales approaching a billion light years — has been an important diagnostic tool in the development of cosmological models and the determination of the mass–energy content and geometry of the Universe[1]. The standard observational approach has involved painstaking surveys of individual galaxies over large cosmic volumes, culminating in the Sloan Digitial Sky Survey[2], which has catalogued a million galaxies to lookback times of a few billion years (or redshifts about 0.3). On page 463 of this issue, Chang *et al.*[3] pioneer a different approach to the study of large-scale structure, termed 'intensity mapping'.

Using low-resolution, low-frequency observations, the authors[3] detect the summed 21-centimetre radio emission by neutral atomic hydrogen (HI) from aggregates of thousands of galaxies at large lookback times (redshift about 1). By smoothing this radio emission over large cosmic volumes, they bypass the need to detect individual galaxies, and determine the large-scale structure directly from the smoothed signal. Although the result is not quite ready for 'prime-time' cosmology, the demonstration of this technique has great promise for future studies of the evolution of large-scale structure with cosmic epoch.

The study of the evolution of large-scale structure has attained new relevance in the past few years, for two reasons. The first is the discovery, using distant stellar explosions called supernovae as standard candles, that the mass–energy content of the Universe since redshift around 1 is dominated by a mysterious 'dark energy', which is thought to accelerate cosmic expansion[4]. And the second is the detection of the primordial 'Baryon Acoustic Oscillations' (BAO).

The BAO result from sounds waves that propagated through the hot plasma of photons, protons, and electrons in the early Universe (redshift $\geq 1000$), and leave their signature in temperature maps of the cosmic microwave background radiation, the relic photons from the Big Bang[5]. The Sloan survey has detected the echo of these sounds waves imprinted on the galaxy distribution in the nearby Universe[6] on a scale of 100 million parsecs. The BAO, being of known intrinsic size, provide a 'standard ruler' that can be used to study the change in cosmic geometry over billions of years. Such measurements then dictate the evolution of cosmic expansion, and hence the nature of dark energy which drives this expansion[7]. Unfortunately, current galaxy surveys are insufficient to probe the BAO to the distances required to address this fundamental question in cosmology. Chang and colleagues' approach[3] offers an alternative which has the potential for tracing the BAO to unprecedented lookback times.

The authors employ the sensitive Green Bank Telescope to map the 21-cm light emitted by HI in distant galaxies; this radiation is released by the atom when it undergoes the

'hyperfine' energy transition. The combined spatial and spectral resolution of this HI 'intensity map' corresponds to a volumetric resolution element of about 40 cubic Mpc. On its own, the current map has insufficient sensitivity to search for large-scale structures directly, let alone to detect the BAO. However, by cross correlating the HI map with the DEEP2 optical galaxy redshift survey[8], the authors are able to detect the summed HI 21-cm signal from optically selected galaxies in the large cosmic volume probed. In this way, they derive a mean atomic neutral gas content at redshift 1 — which applies at least to galaxies selected by their optical emission. This, in itself, has interesting astrophysical implications. But perhaps the most important result from their study is the development of the technique of intensity mapping itself.

Although the method is, in principle, straightforward, there are a number of daunting observational challenges. The first is the strong broadband (continuum) emission that permeates the sky at low radio frequencies. This emission comes both from our Galaxy, in the form of a smooth spatial component, as well as from myriad extragalactic point sources (galaxies and quasars), and it is many orders of magnitude stronger than the expected HI 21-cm signal. Chang and colleagues[3] are able to remove it successfully from their HI maps by using a clever technique known as non-parametric single value decomposition. The second challenge is radio (mostly man-made) interference, for which the authors employ filtering techniques based on the polarization properties of terrestrial signals. Overall, their techniques provide hope that future observations will be able to reach the required sensitivity to map the large-scale structure, and detect the BAO, directly with HI 21-cm intensity mapping.

Beyond cosmology, there is astrophysical interest in determining the evolution of the neutral gas content of galaxies. In essence, galaxy formation (to an astrophysicist) entails the conversion of gas to stars over cosmic time. The most naïve assumption is that, at some point in the past, galaxies were mostly gas, and this gas then fuels star formation. But observations of this phenomenon have presented a puzzle. It is now well quantified that the cosmic star-formation rate density a few billion years ago was an order of magnitude larger than today. In effect, we live in a relatively boring cosmic epoch, and things promise to become more boring with time. However, indirect measurements of the evolution of the cosmic HI mass density, via studies of HI Ly-alpha absorption lines in the spectra of quasars and galaxies, shows essentially no change in the HI mass density over this same cosmic time range, and beyond[9]. This suggests that the gas collects in mostly molecular form ($H_2$), or that the neutral atomic gas is simply seen during a phase transition as it accretes onto galaxies from the ionized intergalactic medium (or some combination thereof[10]).

Chang and colleagues' intensity mapping technique, coupled with surveys of optical galaxies, provides an alternative means of measuring the mean HI mass density in the distant Universe. Their result[3] represents an independent confirmation of the Ly-alpha absorption measurements, supporting the conclusion that the cosmic HI mass density is roughly constant with redshift.

The detection of neutral hydrogen in galaxies at large cosmic distances has been a major science driver for the future Square Kilometre Array (SKA) radio telescope. Indeed, the measurement of the BAO via large surveys of HI 21-cm radio emission from distant galaxies is one of the key science projects for the SKA[11]. Chang and colleagues[3] demonstrate a technique that could provide the first insight into large-scale structure at high redshifts before the construction of such a mega-facility.

Chris L. Carilli works at the Pete V. Domenici Array Science Center, PO Box O, Socorro, New Mexico, USA 87801.

e-mail: ccarilli@nrao.edu

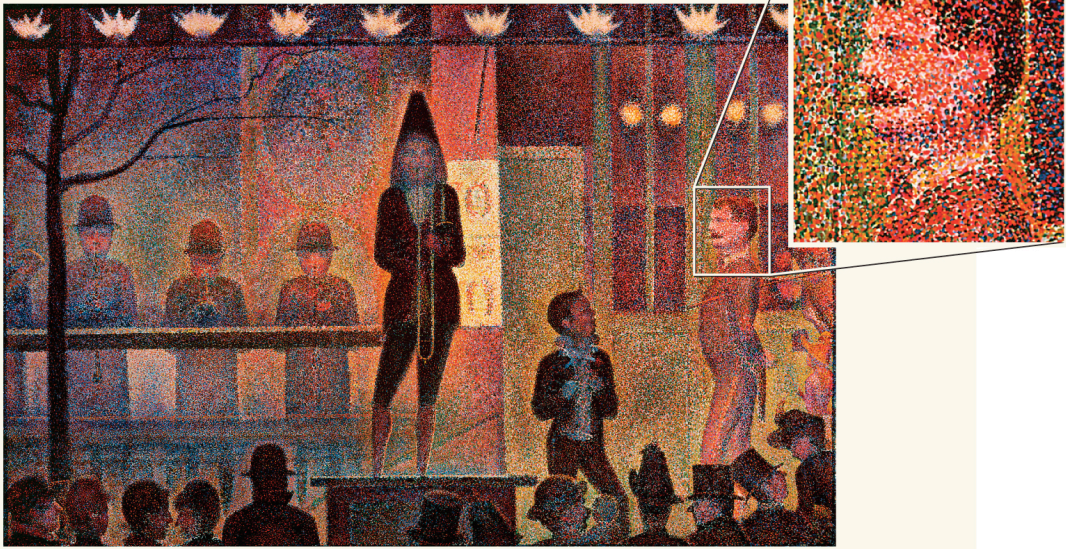

**Figure 1: Cosmic surveys and pointillism.** Traditionally, astronomers map out the large-scale structure in the Universe by observing and cataloguing millions of galaxies — much like a painter employing the technique of pointillism, here depicted in Georges Seurat's painting *La Parade,* uses many small distinct dots to generate an image. Chang and colleagues' survey of 21-cm radio emission by neutral atomic hydrogen from aggregates of thousands of galaxies sidesteps the need to detect the individual sources and looks for large-scale patterns directly — somewhat like using a broad brush to produce a painting of the cosmos.